\documentclass[useAMS,usenatbib,color]{mn2e}
\usepackage{natbib}
\usepackage{amssymb,amsmath,amsfonts}
\usepackage{epsfig}

\title{Probing the Formation of the First Low-Mass Stars with Stellar Archaeology}
\author[A. Frebel, J.L. Johnson and V. Bromm]
       {Anna Frebel\thanks{E-mail: anna@astro.as.utexas.edu}, Jarrett L. Johnson and Volker Bromm \\
McDonald Observatory and Department of Astronomy, University of Texas, Austin, TX 78712, USA \\}

\pubyear{2007}

\begin{document}
\maketitle
\topmargin-1cm

\label{firstpage}

\begin{abstract}
We investigate the conditions under which the first low-mass stars
formed in the Universe by confronting theoretical predictions
governing the transition from massive Population\,III to low-mass
Population\,II stars with recent observational C and/or O abundance
data of metal-poor Galactic stars. We
introduce a new ``observer-friendly'' function, the transition
discriminant $D_{\rm trans}$, which provides empirical constraints as
well as a powerful comparison between the currently available data of
metal-poor halo stars and theoretical predictions of the formation of
the first low-mass stars ($\la 1 M_{\odot}$).  Specifically, we
compare the empirical stellar results with the theory that
fine-structure lines of C and O dominate the transition from Pop\,III
to Pop\,II in the early Universe. We find the currently-available data for
halo stars as well as for dSph galaxies and globular clusters to be
consistent with this theory. An explanation for the observed lack of
metal-poor stars in dSph galaxies and globular clusters is also
suggested. Finally, we predict that any star to be found with
$\mbox{[Fe/H]}\lesssim-4$ should have enhanced C and/or O
abundances. The high C and O abundances of the two most iron-poor
stars are in line with our prediction.
\end{abstract}


\section{Introduction}

One of the most intriguing aspects of modern cosmology is the
formation of the first stars and galaxies. It is now widely believed
that these first stars were very massive; numerical simulations show
that the collapse and fragmentation of primordial gas, where cooling
relies on molecular hydrogen only, leads to Pop\,III stars with masses
$\geq 100 M_{\odot}$ (e.g. \citealt{abel_sci, bromm02, yoshida06}).
With few cooling mechanisms available, low mass star formation appears
to have been initially suppressed. To enable the formation of low mass
($\la 1 M_{\odot}$) Pop\,II stars, additional metal coolants,
generated in the first supernovae, are required (e.g.
\citealt{omukai00, bromm01}). Little is known about
the initial chemical enrichment of the Universe. However, the onset of metal-pollution in the early
Universe must have played an important part in the transition from
massive Pop\,III to low-mass Pop\,II objects observable today.

Currently, two general classes of competing models for the Pop\,III --
Pop\,II transition are discussed in the literature: {\it (i)} atomic fine-structure line
cooling (\citealt{brommnature, santoro06}); and {\it (ii)}
dust-induced fragmentation (e.g. \citealt{schneider06}).  Within the
fine-structure model, C\,II and O\,I have been suggested as main
coolants \citep{brommnature}, such that low-mass star formation can
occur in gas that is enriched beyond critical abundances of
$\mbox{[C/H]}_{\rm crit}\simeq -3.5\pm 0.1$ and $\mbox{[O/H]}_{\rm
crit}\simeq -3 \pm 0.2$.\footnote{$\mbox{[X/Y]} = \log_{10}({N_{\rm
X}/N_{\rm Y}})_{\star}-\log_{10}({N_{\rm X}/N_{\rm Y}})_{\odot}$, for
elements X, Y.} The dust-cooling model, on the other hand, predicts
critical abundances that are typically smaller by a factor of
$10-100$. This order-of-magnitude difference is much larger than the
uncertainties within the fine-structure line model itself, which are
still substantial, regarding the precise value of the critical
abundances. Also, elements other than C and O (such as Si and Fe)
might contribute to the cooling (e.g. \citealt{santoro06}), but are
unlikely to exceed a factor of a few in the overall cooling rate.

Without metals, the primordial gas converges to the characteristic, or
`loitering', state of $T_{\rm char}\simeq 200$\,K and $n_{\rm
char}\simeq 10^{4}$\,cm$^{-3}$. Numerical simulations identify this state
as the main reason for the high mass-scale of Pop\,III stars, because
the gas undergoes a phase of slow, quasi-hydrostatic contraction
during which any inhomogeneities that could seed further
subfragmentation are wiped out by pressure forces (see
\citealt{brommARAA} for details). Cooling from dust, rather than
metals in the gas phase, would only become important at higher
densities, without influencing the evolution toward the loitering state, and
hence the high mass-scale of Pop\,III
star formation (\citet{brommnature}).

Ultimately, this theoretical debate has to be decided empirically.  A
way forward is to extensively test the fine-structure line model with
observational data of C and O abundances found in old Galactic
metal-poor stars.  In order to find the oldest, most ``primordial''
low mass stars, \citet{brommnature} suggested to search for the most
C- and O-poor objects. Such abundances should reflect the level of a
star's primordiality independent of the abundances of heavier elements
(e.g. Fe). They may thus provide an important and feasible
observational tool to constrain the formation mechanisms of the first
low-mass stars in the high-redshift Universe (e.g.
\citealt{salvadori07}) while also offering an empirical discrimination
between the two cooling models. Such constraints, in turn, should also
provide clues to the relevance of the first generations of stars to
cosmic reionization (e.g. \citealt{TVS04}).  Detailed knowledge about
the first epoch of star formation will also allow improved predictions
for future high-redshift observations (e.g. \citealt{BarL07}).
Together with the currently available data, the imminent arrival of
large samples of metal-poor stars such as from SEGUE or LAMOST
\citep{ARAA}, and the Southern Sky Survey in the near future, provides
an ideal test bed to empirically constrain and possibly differentiate
between different cooling theories.

\section{The Transition Discriminant}

To facilitate this test, we first note that we need a criterion for
low-mass star formation that combines cooling due to C\,II and
O\,I. Our discussion is based on the theory
presented in \citet{brommnature}, and we here only summarize the main
points of that theory in an idealized and simplified way for the
convenience of the reader.
To this extent, let us
consider the balance between fine-structure line cooling and adiabatic
compressional heating:
\begin{equation}
\Lambda_{\rm tot}=\Lambda_{\rm C\,II} + \Lambda_{\rm O\,I}\ga \Gamma_{\rm ad}
\mbox{\ ,}
\end{equation}
where all terms have to be evaluated at $n_{\rm char}$, $T_{\rm char}$.
Heating due to adiabatic compression is given by
\begin{equation}
\Gamma_{\rm ad}\simeq 1.5 n_{\rm char}\frac{k_{\rm B}T_{\rm char}}{t_{\rm ff}}
\simeq 2\times 10^{-23}\mbox{ ergs s$^{-1}$ cm$^{-3}$}
\mbox{\ ,}
\end{equation}
where $t_{\rm ff}\sim 5\times 10^5$\,yr is the free-fall time at the
characteristic state.  The cooling terms are (e.g.
\citealt{stahler05}):
\begin{equation}
\Lambda_{\rm O\,I}\simeq
2\times 10^{-20}\mbox{ ergs s$^{-1}$ cm$^{-3}$}\left(\frac{n_{\rm O}}{n_{\rm H}}\right)/
\left(\frac{n_{\rm O}}{n_{\rm H}}\right)_{\odot}\mbox{\ ,}
\end{equation}
and
\begin{equation}
\Lambda_{\rm C\,II}\simeq
6\times 10^{-20}\mbox{ ergs s$^{-1}$ cm$^{-3}$}\left(\frac{n_{\rm C}}{n_{\rm H}}\right)/
\left(\frac{n_{\rm C}}{n_{\rm H}}\right)_{\odot}\mbox{\ .}
\end{equation}
Equation~(1) can then be written as
\begin{equation}
10^{\scriptsize \mbox{[C/H]}} + 0.3\times 10^{\scriptsize \mbox{[O/H]}}\ga 10^{-3.5}
\mbox{\ .}
\end{equation}

We thus introduce a new function, the `transition discriminant':
 \begin{displaymath}
 D_{\rm trans}\equiv \log_{10}\left(
 10^{\scriptsize \mbox{[C/H]}} + 0.3\times 10^{\scriptsize \mbox{[O/H]}}\right)\mbox{\ ,}
 \end{displaymath}

such that low-mass star formation requires $D_{\rm trans}>D_{\rm
trans, crit} \simeq -3.5\pm 0.2$. This function reproduces the
critical values in \citet{brommnature} for the cases where only C or O
are contributing to the cooling at the loitering state, and
approximates the case where both are present by assuming that the
individual cooling rates are simply added. The theoretical uncertainty
estimated here again approximately reflects the analysis in
\citet{brommnature}. However, the true error could be significantly
higher, but is difficult to reliably estimate without carrying out
sophisticated three-dimensional simulations that incorporate all the
relevant microphysics. Such a comprehensive investigation is beyond
the scope of this paper. 

We note here that \citet{brommnature} attempted to test their
fine-structure model with (rather limited) observational data for C
{\it and} O available at the time. The $D_{\rm trans}$ function now
has the important advantage that cosmologically relevant measurements
can be obtained from stellar C and/or O abundances; in particular also
from the common cases where only one of the two abundances is known.

\section{Data on Metal-Poor Stars}
We collected data for C and/or O abundances for a wide range of
metal-poor stars.\footnote{The data table is available on request from
the first author.} In the cases where only one of the two abundances
is available, $D_{\rm trans}$ is a lower limit. Since we are searching
for stars with the $\it{lowest}$ C and O abundances, knowledge of only
one abundance is already sufficient to discriminate whether the star
lies above $D_{\rm trans, crit}$. We are limited, in most cases, to
the observations of metal-poor giants. As these stars ascend the red
giant branch (RGB), internal mixing takes place that changes the
surface C abundance. A consequence of CNO processing is an increase of
the N abundances at the expense of C. Hence, such mixed objects appear
to have low(er) C abundances that do not reflect the chemical
composition with which they were born.

In the absence of information of the N abundance it is difficult to
assess the degree of mixing the star has undergone. A rough estimate,
however, can be obtained from the luminosity of the
objects. \citet{gratton2000} investigated abundances of moderately
metal-poor stars of different evolutionary stages, traced by
luminosity. At $\log L/L_{\odot}\sim2$ they find a decline of C
abundances in line with the star ascending the giant branch beyond the
RGB-bump \citep{charbonnel98} that is responsible for extra mixing
processes.  According to Figure~10 and Table~7 in \citet{gratton2000},
we thus correct the C abundances of stars with luminosities $\log
L/L_{\odot}>2$ by simply applying an appropriate offset.

\citet{spite2005} have available CNO abundances which allowed them to
extensively test their sample of giants for signs of mixing. We
corrected their ``mixed'' giants in the same way as the other giants
with no available ``mixing'' information.  For comparison, we also
calculated C-depletion-corrected lower limits of $D_{\rm trans}$ for
objects that are enhanced in C (but have no O data available). As
expected, the C-rich objects are separated from the ``C-normal''
stars. Amongst those stars are the two most iron-poor stars
\citep{HE0107_Nature,HE1327_Nature} which have large C and O
overabundances.

\begin{figure}
\includegraphics[clip=true,scale=.72]{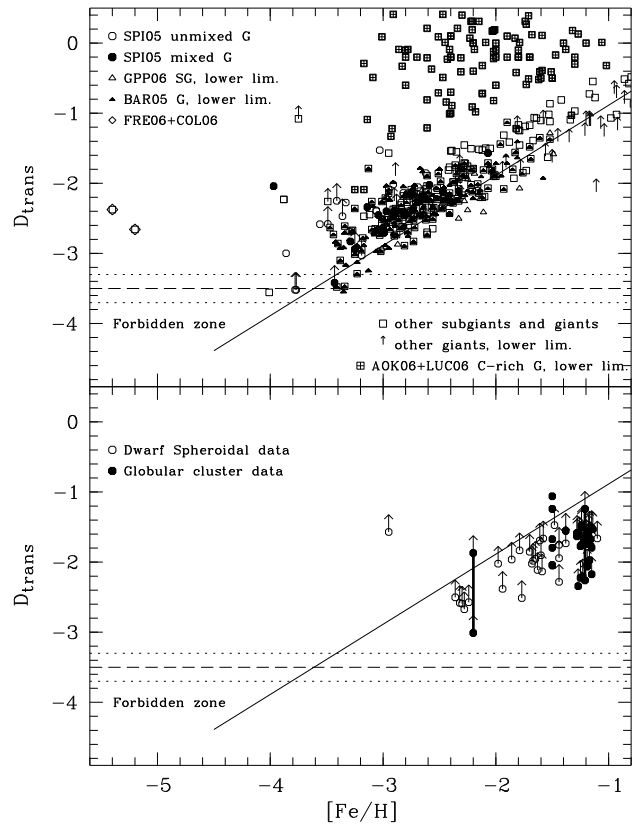}
\caption{\label{dtrans} Transition discriminant, $D_{\rm trans}$, for
 metal-poor stars collected from the literature as a function of
 [Fe/H]. {\it Top panel}: Galactic halo stars.  {\it Bottom panel}:
 Stars in dSph galaxies and globular clusters.  G indicates giants, SG
 subgiants. The critical limit is marked with a dashed line. The
 dotted lines refer to the uncertainty.  References for the data in
 the top panel are SPI06 = \citet{spite2005}, GPP06 =
 \citet{garciaperez_primas2006_O}, BAR05 = \citet{heresII}, AOK06 =
 \citet{aoki_cemp_2006}, LUC06 = \citet{lucatello2006}, FRE06 =
 \citet{o_he1327} and COL06 = \citet{collet06}. ``Other stars'' refer
 to a collection of data from \citet{McWilliametal},
 \citet{israelian98}, \citet{boesgaard99}, \citet{akerman02},
 \citet{aoki_mg}, \citet{carretta02}, \citet{nissen02},
 \citet{bessell05} and \citet{frebel_he1300}. The solid line refers to
 the solar values for C and O \citep{solar_abund}. References for
 bottom panel are \citet{ivans01}, \citet{shetrone01},
 \citet{shetrone03}, \citet{briley2004}, and \citet{fulbright_rich}.
 The solid lines refer to the range of values found in
Cohen, Briley \& Stetson (2002, 2005).}
\end{figure}

In Figure~\ref{dtrans} (top panel) we test $D_{\rm trans, crit}$ with
halo data. We remind the reader that most of the data are lower limits
because of missing C or O abundances. The C-based lower limits,
however, are much more robust than those derived from O because
$D_{\rm trans}$, by definition, is dominated by the C abundance.  All
stars lie above the ``forbidden zone'' (as marked in
Figure~\ref{dtrans}) or within a typical observational error (0.2 to
0.3\,dex) to the critical limit of $D_{\rm trans, crit}$.  There
appears to be a lower envelope of the overall distribution of the
data. Keeping in mind that most of the values are lower limits, this
envelope depicts the solar C and O abundances (solid line).

In the lower panel of Figure~\ref{dtrans} we show globular cluster and
dSph galaxy data collected from the literature. Most of the stars only
have an O abundance available.  With one exception, none of them has a
metallicity less than $\mbox{[Fe/H]}\sim-2.5$. This finding has
recently been investigated by \citet{helmi06} who consistently find a
significant lack of metal-poor stars with $\mbox{[Fe/H]}<-3.0$ in
several dSph galaxies.  In the absence of low metallicity stars these
dSph systems must have had available sufficient amounts of at least C
and O at an early time to avoid the formation of very metal-poor
stars.

\section{Discussion and Implications}

\subsection{Selection Effects}

We caution that the sample of collected data is not complete. The
attempt was rather to populate the diagram with different groups of
objects to study the overall stellar behaviour. Purposely, we included
stars with normal and enhanced C. Any currently seen ``substructure''
(e.g. the two groups that appear to be well separated) within the
data results from our limited selection.  Hence, further data should
be added to fully map the diagram to clearly identify the
``observable'' $D_{\rm trans}$ region(s).

Despite our interest in stars with low C and O abundances, we note
that there is a simple observational bias against C and O deficient
objects. Obviously, it is easier to measure high abundances resulting
in strong spectral features rather than weak ones. In particular, it
is difficult, if not impossible, to detect features from elements of
low abundances in warmer, unevolved main-sequence stars. Insufficient
data quality may lead to additional non-detections of elemental lines
of interest. Hence, C and/or O deficient stars, even if known, likely
have available only upper limits, or no such abundance data at all.

For the region $-4\lesssim\mbox{[Fe/H]}\lesssim-3.0$, close to the
``forbidden zone'' (see Figure~\ref{dtrans}), we attempted to collect
as much data as possible to avoid missing any extremely C-O deficient
objects. Since no stars are currently known with
$-5\lesssim\mbox{[Fe/H]}\lesssim-4$ no data are missing from this
region. If C and O indeed played such an important role in the
formation of the first low-mas stars, it is crucial to obtain further
information by means of more objects (particularly main sequence
stars) with C and O deficiencies. These stars can provide the most
accurate picture of the chemical conditions in the early Universe in
terms of the availability of cooling agents. It is not clear at this
point what the lower end of C and O underabundances may be, and the
current data, strictly speaking, only provide an observational upper
limit on the critical level of $D_{\rm trans}$. We therefore encourage
observers to search for more objects with C and O deficiencies to
obtain good measures of $D_{\rm trans}$ as close as possible to the
critical value, or even below it.

Regardless of observational difficulties, it has been suggested that
the most metal-poor objects are more likely to reside in the bulge
rather than in the halo of the Galaxy \citep{diemand}. This may imply
an additional selection effect of our sample.  However, recent work by
\citep{scannapieco} has shown that the fraction of metal-poor stars in
the halo may not differ greatly from that in the bulge (see also
\citealt{brook}), suggesting that the halo abundances shown in
Figure~\ref{dtrans} may indeed be representative of the most
metal-poor stars in the Milky Way (but see \citealt{GB06}).

\subsection{Constraints on Early Low-Mass Star Formation}
We use stellar archaeology to observationally constrain theoretical
ideas concerning the formation of the first low-mass stars.  In
particular, our literature sample of halo stars provides a firm
observational test for the critical metallicity as predicted by
\citet{brommnature}. No star is lying in the forbidden zone. This
behaviour is consistent with fine-structure line cooling of C and O
governing the transition from Pop\,III to Pop\,II stars in
the early Universe. It is also consistent with the theory based on dust
cooling \citep{schneider06}, which predicts a much lower critical
metallicity. However, this latter theory cannot easily explain the
complete absence of any stars with $-3.5 > D_{\rm trans}\ga -6$.

Another point of interest in Figure~\ref{dtrans} concerns the
crossover of the solar abundance line into the forbidden zone. This
happens at $\mbox{[Fe/H]}\sim-4$. The ``C-normal'' stars in the top
panel well follow the solar trend down to this metallicity.  No
metal-poor stars are known with $\mbox{[Fe/H]}<-4$ with the exception
of two objects that display rather exotic, non-solar-scaled, abundance
patterns. Their unusual behaviour is clearly reflected in our
diagram. If they were to have solar-scaled abundances at
$\mbox{[Fe/H]}<-5$, they would need to follow the trend given by all
the other stars and would fall right into the forbidden zone. This
would have been in conflict with our prediction. We note though, that
stars with $D_{\rm trans}< -3.5$ might have formed in the early
Universe. However, according to the \citet{brommnature} theory, they
would not have been of low-mass, and hence would have already died a
long time ago, rendering them unobservable in present-day surveys.  As
a consequence, any stars with iron abundances lower than
$\mbox{[Fe/H]}\sim-4$ should not display scaled-down solar abundance
patterns as already found in the two stars with
$\mbox{[Fe/H]}<-5$. The chemical behaviour, at least concerning C and
O, should be very different, in line with a $D_{\rm trans}$ value
above the critical threshold.

\subsection{Dwarf Galaxy and Globular Cluster Populations}
The paucity of stars with $\mbox{[Fe/H]}<-2.5$ in dSph galaxies and
globular clusters, as compared to those in the Galactic halo, suggests
that the most metal-poor stars in these systems formed in a different
environment (e.g. \citealt{helmi06}). Many globular clusters are
likely to have formed in merging galaxies,
(e.g. \citealt{ashman_zepf92, ashman_zepf01}).  Likewise, it has been
suggested that some dwarf galaxies may form in the tidal tails of
merging galaxies \citep{barnes,elmegreen93, wetzstein}.  Also,
interestingly, the oldest populations of stars in local dwarf galaxies
appear to be very similar in age to the oldest Galactic globular
clusters \citep{grebel04}, consistent with these systems having formed
through mergers in the early evolution of the Milky Way and the Local
Group.

If globular clusters and dwarf galaxies do indeed form in the mergers
of previous-generation galaxies, then the initial metallicity of the
gas in these systems would have been that of the larger, merging
galaxies responsible for their formation. The most metal-poor globular
cluster and dwarf galaxy stars would then have formed from gas
enriched by previous episodes of star formation that occurred in their
parent merging galaxies. The merging galaxies themselves
would, in turn, contain halo stars that formed at earlier epochs
characterised by gas of lower metallicity. This may explain the lack
of stars with $\mbox{[Fe/H]}\lesssim-3$ observed in local dSph galaxies, as
well as the difference between the metal-poor tail of the metallicity
distribution of these systems and that of the Galactic halo
\citep{helmi06, MetzK07}.

Recent simulations of the formation of dwarf galaxies from primordial gas
at high redshift \citep{ricotti, kawata06, ripamonti} predict a metal-poor
tail of the metallicity distribution that extends well below
$\mbox{[Fe/H]}<-3$. Clearly, this is not observed for the nearby dSph
galaxies observed by \citet{helmi06}.  This may be further evidence
that some dwarf galaxies are formed in the mergers of pre-enriched
larger galaxies, as opposed to being formed from primordial gas in the
early Universe.

\subsection{Low-mass Star Formation at High Redshift}
While the enrichment of the primordial gas to levels of $D_{\rm
trans}\ga-3.5$ may be required for the formation of low-mass stars,
this may not be a sufficient requirement on its own.  It has been
suggested that the temperature floor set by the cosmic microwave
background, $T_{\rm CMB} = 2.7{\rm \,K\,}(1 + z)$, may imply a
redshift dependent characteristic mass of stars formed in the early
Universe (e.g. \citealt{uehara, clarke_bromm, johnson_bromm06}; but
see also \citealt{omukai05}). Gas cannot radiatively cool to below
$T_{\rm CMB}$, hence, the fragmentation mass in collapsing gas clouds
may be larger, in general, at higher redshifts.  Thus, low-mass stars
($\la 1 M_{\odot}$) that can survive until today may not readily form
above a certain redshift, $z_{\rm trans}$, above which only higher
mass stars are likely to form, regardless of the metallicity of the
gas.

This scenario suggests an explanation for the observed lack of scatter
in the abundance ratios of homogeneously analysed low metallicity
stars (\citealt{cayrel2004}; see also \citealt{karlsson2005}).  While
the formation of massive stars likely begins at redshifts $z\ga20$,
stars with masses $\la1M_{\odot}$ may not be able to form until later
times when the $T_{\rm CMB}$ is sufficiently low, at redshifts $z\la
z_{\rm trans}$. This delay between the onset of high-mass and low-mass
star formation may have given ample time for the ejecta from many
supernovae to become mixed in the buildup of the Galaxy before
low-mass stars were finally able to form.  If indeed there is an
explicit redshift dependence on the minimum mass of stars that may
form, then future observations of high redshift systems should find
that only massive stars form at redshifts $z\ga z_{\rm trans}$,
largely independent of the degree of metal enrichment in these
systems. The recent discovery of a massive post-starburst galaxy at
$z\sim6.5$ by \citet{mobasher05} allows to place a firm limit of
$z_{\rm trans}\ga 7$.

\section{Conclusion}
Our literature collection of metal-poor abundance data
supports the notion that fine-structure C and O line
cooling predominantly drives the transition from Pop\,III to Pop\,II stars
in the early Universe. To refine or refute this theory, more
abundances of existing and future metal-poor star data should be added
to Figure~\ref{dtrans}. The ultimate goal is to understand the details
of when and how the first low-mass stars formed. Detailed knowledge of
this cosmic milestone will provide insight into the earliest history
of galaxy formation, and the nature of the first stars and supernovae.

\section*{Acknowledgements} We thank Achim Weiss for discussions on RGB
mixing. A.~F. acknowledges support through the W.~J.~McDonald
Fellowship of the McDonald Observatory.



\begin{thebibliography}{60}

\expandafter\ifx\csname natexlab\endcsname\relax\def\natexlab#1{#1}\fi

\bibitem[{{Abel} {et~al.}(2002){Abel}, {Bryan}, {Norman}}]{abel_sci}
{Abel} T., {Bryan} G.~L., {Norman} M.~L., 2002, Sci, 295, 93

\bibitem[{{Akerman} {et~al.}(2004){Akerman}, {Carigi}, {Nissen}, {Pettini},
  {Asplund}}]{akerman02}
{Akerman} C.~J., {Carigi} L., {Nissen} P.~E., {Pettini} M., {Asplund}
  M., 2004, A\&A, 414, 931

\bibitem[{{Aoki} {et~al.}(2006){Aoki}, {Beers}, {Christlieb}, {Norris}, {Ryan},
  {Tsangarides}}]{aoki_cemp_2006}
{Aoki} W. et al., 2006, ApJ, submitted

\bibitem[{{Aoki} {et~al.}(2002){Aoki}, {Norris}, {Ryan}, {Beers},
  {Ando}}]{aoki_mg}
{Aoki} W., {Norris} J.~E., {Ryan} S.~G., {Beers} T.~C., {Ando} H., 2002,
  ApJL, 576, L141

\bibitem[{{Ashman} \& {Zepf}(1992)}]{ashman_zepf92}
{Ashman} K.~M., {Zepf} S.~E., 1992, ApJ, 384, 50

\bibitem[{{Ashman} \& {Zepf}(2001)}]{ashman_zepf01}
{Ashman} K.~M., {Zepf} S.~E., 2001, AJ, 122, 1888

\bibitem[{{Asplund} {et~al.}(2005){Asplund}, {Grevesse},
  {Sauval}}]{solar_abund}
{Asplund} M., {Grevesse} N., {Sauval} A.~J., 2005, in ASP Conf. Ser. 336:
  Cosmic Abundances as Records of Stellar Evolution and Nucleosynthesis, 25

\bibitem[{{Barkana}\& {Loeb}(2007)}]{BarL07}
{Barkana} R., {Loeb} A., 2007, Rep. Prog. Phys., in press (astro-ph/0611541)

\bibitem[{{Barklem} {et~al.}(2005){Barklem}, {Christlieb}, {Beers}, {Hill},
  {Bessell}, {Holmberg}, {Marsteller}, {Rossi}, {Zickgraf},
  {Reimers}}]{heresII}
{Barklem} P.~S. et al., 2005, A\&A, 439, 129

\bibitem[{{Barnes} \& {Hernquist}(1992)}]{barnes}
{Barnes} J.~E., {Hernquist} L., 1992, Nat, 360, 715

\bibitem[{{Beers} \& {Christlieb}(2005)}]{ARAA}
{Beers} T.~C., {Christlieb} N., 2005, ARA\&A, 43, 531

\bibitem[{{Bessell} \& {Christlieb}(2005)}]{bessell05}
{Bessell} M.~S. {Christlieb} N., 2005, in IAU Symposium, ed. V.~{Hill},
  P.~{Fran{\c c}ois}, F.~{Primas}, 237

\bibitem[{{Boesgaard} {et~al.}(1999){Boesgaard}, {King}, {Deliyannis},
  {Vogt}}]{boesgaard99}
{Boesgaard} A.~M., {King} J.~R., {Deliyannis} C.~P., {Vogt} S.~S., 1999,
  AJ, 117, 492

\bibitem[{{Briley}, {Cohen} \& {Stetson}(2004)}] {briley2004}
{Briley} M.~M., {Cohen} J.~G., {Stetson} P.~B, 2004, AJ, 127, 1579

\bibitem[{{Bromm} {et~al.}(2002){Bromm}, {Coppi}, {Larson}}]{bromm02}
{Bromm} V., {Coppi} P.~S., {Larson} R.~B., 2002, ApJ, 564, 23

\bibitem[{{Bromm} {et~al.}(2001){Bromm}, {Ferrara}, {Coppi},
  {Larson}}]{bromm01}
{Bromm} V., {Ferrara} A., {Coppi} P.~S., {Larson} R.~B., 2001, MNRAS,
  328, 969

\bibitem[{{Bromm} \& {Larson}(2004)}]{brommARAA}
{Bromm} V., {Larson} R.~B., 2004, ARA\&A, 42, 79

\bibitem[{{Bromm} \& {Loeb}(2003)}]{brommnature}
{Bromm} V., {Loeb} A., 2003, Nat, 425, 812

\bibitem[{{Brook} {et~al.}(2006){Brook}, {Kawata}, {Scannapieco}, {Martel},
  {Gibson}}]{brook}
{Brook} C.~B., {Kawata} D., {Scannapieco} E., {Martel} H., {Gibson}
  B.~K., 2006, ApJ, in press (astro-ph/0612259)

\bibitem[{{Carretta} {et~al.}(2002){Carretta}, {Gratton}, {Cohen}, {Beers},
  {Christlieb}}]{carretta02}
{Carretta} E., {Gratton} R., {Cohen} J.~G., {Beers} T.~C., {Christlieb}
  N., 2002, A\&A, 124, 481

\bibitem[{{Cayrel} {et~al.}(2004){Cayrel}, {Depagne}, {Spite}, {Hill}, {Spite},
  {Fran{\c c}ois}, {Plez}, {Beers}, {Primas}, {Andersen}, {Barbuy},
  {Bonifacio}, {Molaro}, {Nordstr{\"o}m}}]{cayrel2004}
{Cayrel} R. et al., 2004, A\&A, 416,
  1117

\bibitem[{{Charbonnel} {et~al.}(1998){Charbonnel}, {Brown},
  {Wallerstein}}]{charbonnel98}
{Charbonnel} C., {Brown} J.~A., {Wallerstein} G., 1998, A\&A, 332, 204

\bibitem[{{Christlieb} {et~al.}(2002){Christlieb}, {Bessell}, {Beers},
  {Gustafsson}, {Korn}, {Barklem}, {Karlsson}, {Mizuno-Wiedner},
  {Rossi}}]{HE0107_Nature}
{Christlieb} N. et al., 2002, Nat, 419, 904

\bibitem[{{Clarke} \& {Bromm}(2003)}]{clarke_bromm}
{Clarke} C.~J., {Bromm} V., 2003, MNRAS, 343, 1224

\bibitem[{{Cohen}, {Briley} \& {Stetson}(2005)}]{cohen2005}
{Cohen} J.~G., {Briley} M.~M., {Stetson} P.~B., 2005, AJ, 130, 1177

\bibitem[{{Cohen}, {Briley} \& {Stetson}(2002)}] {cohen2002}
{Cohen} J.~G., {Briley} M.~M., {Stetson} P.~B., 2002, AJ, 123, 2525

\bibitem[{{Collet} {et~al.}(2006){Collet}, {Asplund},
  {Trampedach}}]{collet06}
{Collet} R., {Asplund} M., {Trampedach} R., 2006, ApJL, 644, L121

\bibitem[{{Diemand} {et~al.}(2005){Diemand}, {Madau}, {Moore}}]{diemand}
{Diemand} J., {Madau} P., {Moore} B., 2005, MNRAS, 364, 367

\bibitem[{{Elmegreen} {et~al.}(1993){Elmegreen}, {Kaufman},
  {Thomasson}}]{elmegreen93}
{Elmegreen} B.~G., {Kaufman} M., {Thomasson} M., 1993, ApJ, 412, 90

\bibitem[{{Frebel} {et~al.}(2005){Frebel}, {Aoki}, {Christlieb}, {Ando},
  {Asplund}, {Barklem}, {Beers}, {Eriksson}, {Fechner}, {Fujimoto}, {Honda},
  {Kajino}, {Minezaki}, {Nomoto}, {Norris}, {Ryan}, {Takada-Hidai},
  {Tsangarides}, {Yoshii}}]{HE1327_Nature}
{Frebel} A. et al., 2005, Nat, 434, 871

\bibitem[{{Frebel} {et~al.}(2006{\natexlab{a}}){Frebel}, {Christlieb},
  {Norris}, {Aoki}, {Asplund}}]{o_he1327}
{Frebel} A., {Christlieb} N., {Norris} J.~E., {Aoki} W., {Asplund}, M.
  2006{\natexlab{a}}, ApJL, 638, L17

\bibitem[{{Frebel} {et~al.}(2006{\natexlab{b}}){Frebel}, {Norris}, {Aoki},
  {Honda}, {Bessell}, {Takada-Hidai}, {Beers}, {Christlieb}}]{frebel_he1300}
{Frebel} A. et al., 2006{\natexlab{b}}, ApJ, 658, 534

\bibitem[{{Fulbright} {et~al.}(2004){Fulbright}, {Rich},
  {Castro}}]{fulbright_rich}
{Fulbright} J.~P., {Rich} R.~M., {Castro} S., 2004, ApJ, 612, 447

\bibitem[{{Garc{\'{\i}}a P{\'e}rez} {et~al.}(2006){Garc{\'{\i}}a P{\'e}rez},
  {Asplund}, {Primas}, {Nissen}, {Gustafsson}}]{garciaperez_primas2006_O}
{Garc{\'{\i}}a P{\'e}rez} A.~E., {Asplund} M., {Primas} F., {Nissen} P.~E.,
  {Gustafsson} B., 2006, A\&A, 451, 621

\bibitem[{{Gratton} {et~al.}(2000){Gratton}, {Sneden}, {Carretta},
  {Bragaglia}}]{gratton2000}
{Gratton} R.~G., {Sneden} C., {Carretta} E., {Bragaglia} A., 2000, A\&A,
  354, 169

\bibitem[{{Grebel} \& {Gallagher}(2004)}]{grebel04}
{Grebel} E.~K. {Gallagher} III, J.~S., 2004, ApJL, 610, L89

\bibitem[{{Greif} \& {Bromm}(2006)}]{GB06}
{Greif} T.~H. {Bromm} V., 2006, MNRAS, 373, 128

\bibitem[{{Helmi} {et~al.}(2006){Helmi}, {Irwin}, {Tolstoy}, {Battaglia},
  {Hill}, {Jablonka}, {Venn}, {Shetrone}, {Letarte}, {Arimoto}, {Abel},
  {Francois}, {Kaufer}, {Primas}, {Sadakane}, {Szeifert}}]{helmi06}
{Helmi} A. et al., 2006, ApJL, 651, L121

\bibitem[{{Israelian} {et~al.}(1998){Israelian}, {Garc{\'{\i}}a L{\'o}pez},
  {Rebolo}}]{israelian98}
{Israelian} G., {Garc{\'{\i}}a L{\'o}pez} R.~J., {Rebolo} R., 1998, ApJ,
  507, 805

\bibitem[{{Ivans} {et~al.}(2001){Ivans}, {Kraft}, {Sneden}, {Smith}, {Rich},
  {Shetrone}}]{ivans01}
{Ivans} I.~I. et al., 2001, AJ, 122, 1438

\bibitem[{{Johnson} \& {Bromm}(2006)}]{johnson_bromm06}
{Johnson} J.~L., {Bromm} V., 2006, MNRAS, 366, 247

\bibitem[{{Karlsson} \& {Gustafsson}(2005)}]{karlsson2005}
{Karlsson} T., {Gustafsson} B., 2005, A\&A, 436, 879

\bibitem[{{Kawata} {et~al.}(2006){Kawata}, {Arimoto}, {Cen},
  {Gibson}}]{kawata06}
{Kawata} D., {Arimoto} N., {Cen} R., {Gibson} B.~K., 2006, ApJ, 641, 785

\bibitem[{{Lucatello} {et~al.}(2006)}]{lucatello2006}
{Lucatello} S. et al., 2006, ApJ, submitted

\bibitem[{{McWilliam} {et~al.}(1995){McWilliam}, {Preston}, {Sneden},
  {Searle}}]{McWilliametal}
{McWilliam} A., {Preston} G.~W., {Sneden} C., {Searle} L., 1995, AJ, 109,
  2757

\bibitem[{{Metz} \& {Kroupa}(2007)}]{MetzK07}
{Metz} M., {Kroupa} P., 2007, MNRAS, 376, 387

\bibitem[{{Mobasher} {et~al.}(2005){Mobasher}, {Dickinson}, {Ferguson},
  {Giavalisco}, {Wiklind}, {Stark}, {Ellis}, {Fall}, {Grogin}, {Moustakas},
  {Panagia}, {Sosey}, {Stiavelli}, {Bergeron}, {Casertano}, {Ingraham},
  {Koekemoer}, {Labbe'}, {Livio}, {Rodgers}, {Scarlata}, {Vernet}, {Renzini},
  {Rosati}, {Kuntschner}, {Kummel}, {Walsh}, {Chary}, {Eisenhardt},
  {Stern}}]{mobasher05}
{Mobasher} B. et al.,  2005, ApJ, 635, 832

\bibitem[{{Nissen} {et~al.}(2002){Nissen}, {Primas}, {Asplund},
  {Lambert}}]{nissen02}
{Nissen} P.~E., {Primas} F., {Asplund} M., {Lambert} D.~L., 2002, A\&A,
  390, 235

\bibitem[{{Omukai}(2000)}]{omukai00}
{Omukai} K., 2000, ApJ, 534, 809

\bibitem[{{Omukai} {et~al.}(2005){Omukai}, {Tsuribe}, {Schneider},
  {Ferrara}}]{omukai05}
{Omukai} K., {Tsuribe} T., {Schneider} R., {Ferrara} A., 2005, ApJ, 626,
  627

\bibitem[{{Ricotti} \& {Gnedin}(2005)}]{ricotti}
{Ricotti} M., {Gnedin} N.~Y. 2005, ApJ, 629, 259

\bibitem[{{Ripamonti} {et~al.}(2006){Ripamonti}, {Tolstoy}, {Helmi},
  {Battaglia}, {Abel}}]{ripamonti}
{Ripamonti} E., {Tolstoy} E., {Helmi} A., {Battaglia} G., {Abel} T.,
  2006, astro-ph/0612210

\bibitem[{{Salvadori} {et~al.}(2007){Salvadori}, {Schneider}, {Ferrara}}]
{salvadori07} {Salvadori} S., {Schneider} R., {Ferrara} A., 2007,
MNRAS, submitted (astro-ph/0611130)

\bibitem[{{Santoro} \& {Shull}(2006)}]{santoro06}
{Santoro} F., {Shull} J.~M., 2006, ApJ, 643, 26

\bibitem[{{Scannapieco} {et~al.}(2006){Scannapieco}, {Kawata}, {Brook},
  {Schneider}, {Ferrara}, {Gibson}}]{scannapieco}
{Scannapieco} E. et al., 2006, ApJ, 653, 285

\bibitem[{{Schneider} {et~al.}(2006){Schneider}, {Omukai}, {Inoue},
  {Ferrara}}]{schneider06}
{Schneider} R., {Omukai} K., {Inoue} A.~K., {Ferrara} A., 2006, MNRAS,
  369, 1437

\bibitem[{{Shetrone} {et~al.}(2003){Shetrone}, {Venn}, {Tolstoy}, {Primas},
  {Hill}, {Kaufer}}]{shetrone03}
{Shetrone} M. et al., 2003, AJ, 125, 684

\bibitem[{{Shetrone} {et~al.}(2001){Shetrone}, {C{\^o}t{\'e}},
  {Sargent}}]{shetrone01}
{Shetrone} M.~D., {C{\^o}t{\'e}} P., {Sargent} W.~L.~W., 2001, ApJ, 548,
  592

\bibitem[{{Spite} {et~al.}(2005){Spite}, {Cayrel}, {Plez}, {Hill}, {Spite},
  {Depagne}, {Fran{\c c}ois}, {Bonifacio}, {Barbuy}, {Beers}, {Andersen},
  {Molaro}, {Nordstr{\"o}m}, {Primas}}]{spite2005}
{Spite} M. et al., 2005, A\&A, 430, 655

\bibitem[{{Stahler} \& {Palla}(2005)}]{stahler05}
{Stahler} S.~W., {Palla} F., 2005, {The Formation of Stars} (Weinheim:
  Wiley-VCH)

\bibitem[{{Tumlinson} {et~al.}(2004){Tumlinson}, {Venkatesan}, {Shull}}]{TVS04}
{Tumlinson} J., {Venkatesan} A., {Shull} J.~M., 2004, ApJ, 612, 602

\bibitem[{{Wetzstein} {et~al.}(2005){Wetzstein}, {Naab},  {Burkert}}]{wetzstein}
{Wetzstein} M., {Naab} T., {Burkert} A., 2005, MNRAS, 375, 805

\bibitem[{{Uehara}\&  {Inutsuka}(2000)}]{uehara}
{Uehara} H., {Inutsuka} S., 2000, ApJ, 531, L91


\bibitem[{{Yoshida} {et~al.}(2006){Yoshida}, {Omukai}, {Hernquist},
  {Abel}}]{yoshida06}
{Yoshida} N., {Omukai} K., {Hernquist} L., {Abel} T., 2006, ApJ, 652, 6

\end{thebibliography}
\end{document}